\documentclass[12pt,a4paper]{article}
\usepackage[utf8]{inputenc}
\usepackage{amsmath}
\usepackage{amsfonts}
\usepackage{amssymb}
\usepackage{bm}
\usepackage{lscape}
\usepackage{pdflscape}
\usepackage[margin=1in,footskip=0.25in]{geometry}
\usepackage{graphicx}
\usepackage{multirow}
\usepackage[table,xcdraw]{xcolor}

\usepackage{enumerate}
\usepackage{lscape}
\usepackage{url}

\begin{document}

\noindent \textbf{\Large{Changing Clusters of Indian States with respect to number of Cases of COVID-19 using incrementalKMN Method}}\\
\\
Rabinder Kumar Prasad\textsuperscript{a} \,\,\, Rosy Sarmah\textsuperscript{b}\,\,and Subrata Chakraborty\textsuperscript{c}\\
\\
\textsuperscript{a}Department of CSE, 
Dibrugarh University, 
Dibrugarh-786004, 
India\\
\\
\textsuperscript{b}Department of CSE,Tezpur University,
Assam, 784028,
India\\
\\
\textsuperscript{c}Department of Statistics,
Dibrugarh University,
Dibrugarh-786004,
India\\
\\
Email:\,\,\textsuperscript{a}\textit{rkp@dibru.ac.in}\,\,\,\,\,\textsuperscript{c}\textit{rosy8@tezu.ernet.in} \,\,\,\,\,\textsuperscript{c}\textit{subrata\_stats@dibru.ac.in}\\
\\
\section*{Abstract}
The novel Coronavirus (COVID-19) incidence in India is currently experiencing exponential rise but with apparent spatial variation in growth rate and doubling time rate. We classify the states into five clusters with low to the high-risk category and study how the different states moved from one cluster to the other since the onset of the first case on $30^{th}$ January 2020 till the end of unlock 1 that is $30^{th}$ June 2020. We have implemented a new clustering technique called the incrementalKMN (Prasad, R. K., Sarmah, R., Chakraborty, S.(2019)) clustering. 
\section*{keywords}
COVID-19, clustering, growth rate, doubling time.

\section{Introduction}
\label{intro}
A severe acute respiratory disease, caused by a novel coronavirus, (COVID-19) has spread in the month of November- December 2019 throughout China and received worldwide attention. The World Health Organization (WHO) officially declared the novel coronavirus (COVID-19) epidemic on $30^{th}$ January 2020 as a public health emergency of international concern.
\newline
In India, the first case of novel coronavirus (COVID-19) was detected on $30^{th}$ January in the State of Kerala (Ward, A. (2020, March 24)). As the number of confirmed novel coronavirus positive cases closed 500, the Govt. of India introduced ``Janta Curfew" on $19^{th}$ March 2020 and after that Govt. of India enforced 21 days lockdown Phase-I nationwide from $25^{th}$ March – $14^{th}$ April, nearly all services and factories were suspended (Singh, K. D., Goel, V., Kumar, H., Gettleman, J. (2020, March 25)). 
\newline
Due to the number of confirmed cases of COVID-19 increased, on $14^{th}$ April 2020, Govt. of India extended the lockdown Phase-I period till $3^{rd}$ May 2020 i.e. lockdown Phase-II, with certain relaxations (Bhaskar, U. (2020, April 14), Dutta, P. K. (2020, April 14)). In Phase-II, the lockdown area were categorised into \textit{three} zone ``red zone", ``orange zone", and ``green zone" based on number of confirmed cases (BBC, 2020, April 16).
\newline
On $1^{st}$ May 2020, the Ministry of Home Affairs (MHA) and the Government of India (GoI) further extended the lockdown period i.e Phase-III to two weeks beyond 4 May based on the number of cases, with some relaxations ( Online, E. T. (2020, May 4), newsworld24. (2020, May 2)).
\newline
Again, the National Disaster Management Authority (NDMA) and the Ministry of Home Affairs (MHA) extended the lockdown i.e. Phase-IV for a period for two weeks from $18^{th}-31^{st}$ May 2020 on $17^{th}$ May 2020 with additional relaxations. In this phase, the local bodies were given the authority to demarcate containment and buffer zones (Banerjea, W. B. A. (2020, May 17), Desk, T. H. N. (2020, May 18) ,Online, E. T. (2020, May 21)).
\newline
Focused on the economy of India, the MHA issued fresh guidelines for the month of June 2020 based on containment zone and known as Phase-V. This phase is also known as ``Unlock 1" (Sharma, N., Ghosh, D. (2020, May 30)).
\newline
Cluster analysis is an important data analysis tool that searches patterns in a data set by grouping the data objects into clusters with respect to certain criteria. It plays a significant role in almost every area of science and engineering, like video processing, image processing, text analysis, bioinformatics, market research, privacy and security, wireless sensor networks, web social networks analysis, and document clustering, etc (Han, Jiawei, Jian Pei, and Micheline Kamber (2011)). 
\newline
The current surge in the number of COVID-19 cases has created a new social-demographic problem across India with a lot of fatalities. The machine learning community is continuously trying to predict the outcome of the disease in terms of number of cases, number of fatalities, etc. and has been quite successful at that. One can see  (Zarikas V, Poulopoulos SG, Gareiou Z, Zervas E (2020), Kumar S (2020) ,Kumar A, Gupta PK, Srivastava A (2020) ,Blumenstock J (2020) ,Ozturk T, Talo M, Yildirim EA, et al (2020) ,Javaid M, Haleem A, Vaishya R, et al (2020) ,Hassanien AE, Mahdy LN, Ezzat KA, et al (2020)) for different approaches in the literature. 
\newline
In this article, our main objective is to classify the states and UTs of India with respect to the daily incidence of positive cases of COVID-19 in five clusters designated as low to high-risk category. Then to study the dynamics of how the states and UTs changed clusters in phases over the period stretching from $1^{st}$ case till then lockdowns and up to the end of the first unlock period that is $30^{th}$ June 2020. For the process of clustering, we have used the Incremental \textit{k}-means clustering(incrementalKMN) method (Prasad, R. K., Sarmah, R., Chakraborty, S.,(2019)) which is based on \textit{k}-means (Han, Jiawei, Jian Pei, and Micheline Kamber (2011)) clustering method and provides improvement over the later in the quality of clusters in term of minimizing the total Sum of Squared Error (SSE) (Han, Jiawei, Jian Pei, and Micheline Kamber (2011)).

\section{Methodology}
\subsection{Clustering}
In order to define data clustering, let $\textbf{D} =  \{ x_{1} , x_{2} ,……., x_{n} \} $, be a data set with $n$ number of data elements, and each data element characterized with \textit{m} number of features : \textbf{$x_{i}$}= \{$ x_{i,1},x_{i,2},……..x_{i,m}$ \}.  The main objective of clustering is to group these data elements into homogeneous sub-groups such that the intra-cluster similarities are high while inter-cluster similarities are low. The data elements in each sub-group are called a cluster, and the union of all sub-groups is equal to the dataset \textbf{D}. Clustering methods have been classified into five different categories, i.e., Partition-based, density-based, hierarchical-based, grid-based and model-Based (Han, Jiawei, Jian Pei, and Micheline Kamber (2011)). 
In the last few decades, clustering algorithms have been extensively used to solve the problem of data-mining. 
\newline
In this research, we have used incrementalKMN (Prasad, R. K., Sarmah, R., Chakraborty, S.,(2019)) clustering method on novel coronavirus (COVID-19) data set of India based on confirmed cases, which produced the \textit{k} desired number of group of states of India i.e. divide the data set into \textit{k} number of group of states. And apply the growth rate and doubling time from equation \ref{eq1} and equation \ref{eq2} on desired \textit{k} number of clusters produced by incrementalKMN (Prasad, R. K., Sarmah, R., Chakraborty, S.,(2019)) clustering method. The steps of incrementalKMN method is given below:
\begin{enumerate}[Step 1:]
	\item Select the value of $\textit{k}$ and dataset $D$
	\item Set $i=1$ and $C=\phi$, where $C$ is empty centroid list. 
	\item Select first centroid i.e. $c_{i}$ as mean of a given dataset $D$.
	\item Update the centroid list $C=C \cup \{c_{i}\}$.
	\item Assign each data objects to its nearest centroid.
	\item Compute the $SSE$ of each cluster.
	\item Select $i^{th}$ center i.e. $(i \in { 2,3,...,k})$ is selected from maximum $SSE$ of a cluster. The $i^{th}$ center is a maximum distance from the data object and the centroid of maximum $SSE$ cluster.
	\item  Repeat the step $4$ until it reaches \textit{k} number of centroids and finally formed \textit{k} number of clusters.
\end{enumerate}

\subsection{Compound Growth Rate (Murphy, C. B. (2020, May 15)):}
In this paper, we have used compound growth rate over regular time intervals of confirmed case for each state phase wise. The growth rate of each state is computed as:
\begin{equation}
Growth\; rate=\left(\dfrac{Present\_Confirmed\_Case}{Past\_Confirmed\_Case}\right)^{\left(\frac{1}{n}\right)} -1
\label{eq1}
\end{equation}
where, $Present\_Confirmed\_Case$ is a ending value, $Past\_Confirmed\_Case$ is a beginning value and $n$ is a the number of periods(in days). 

\subsection{Doubling Time (Manias, M. (2020, January 10)):}
\begin{figure}
	\centering
	\includegraphics[width=8cm,height=8cm,keepaspectratio]{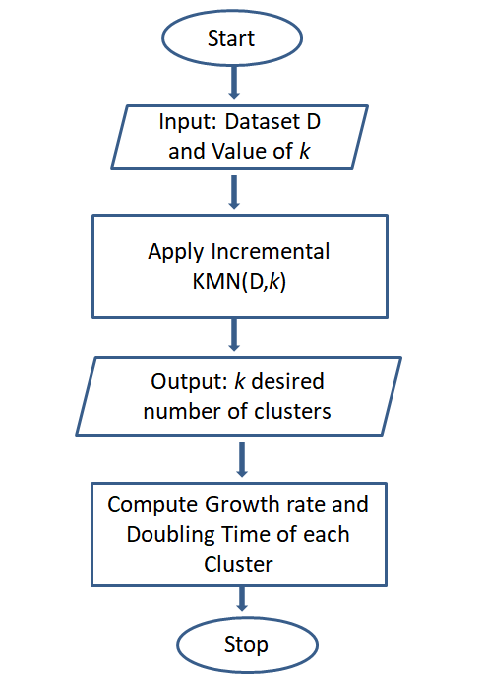}
	\caption{The flowchart of Proposed Method}
	\label{fig7}
\end{figure}

Doubling time is a time it takes for a confirmed case to double in size. The doubling time of confirmed case for each cluster is computed with the help of equation \ref{eq1}, which is described as:  
\begin{equation}
Doubling\;Time= \dfrac{ln(2)}{ln(1+Growth\;rate)}
\label{eq2}
\end{equation}
where, $ln$ stand for natural logarithm.
\newline
The  complete flowchart of the methodology adopted is given in Fig \ref{fig7}. 
\newline
In Fig \ref{fig7}, the proposed method start with input data set $D$ and value of \textit{k}, in this paper, we have considered the value of \textit{k} is \textit{five}. In the next step, we have applied the incrementalKMN clustering method using data set $D$ and value of \textit{k} that produced \textit{k} number of clusters. In the next phase of the flowchart, it computes the growth rate and doubling time of clusters produced by the incrementalKMN clustering method.

\section{Data Source}
The data set of novel coronavirus  (COVID-19) daily confirmed cases state-wise is collected form  (\url{ https://api.covid19india.org/}).In this data set, the total number of confirmed cases in India was 4,14,970 from $30^{th}$ January to $20^{th}$ June 2020. In novel coronavirus dataset, we have considered the following states and 
UTs, as AN(Andaman and Nicobar Islands), AP(Andhra Pradesh), AR (Arunachal Pradesh), AS (Assam), BR (Bihar), CH (Chandigarh), CT (Chhattisgarh), DN (Dadra and Nagar Haveli and Daman and Diu), DL(Delhi), GA (Goa), GJ (Gujarat), HR (Haryana), HP (Himachal Pradesh), JK (Jammu and Kashmir), JH (Jharkhand), KA (Karnataka), KL (Kerala), LA (Ladakh), MP (Madhya Pradesh), MH (Maharashtra), MN (Manipur), ML(Meghalaya), MZ(Mizoram), NL (Nagaland), OR (Odisha), PY (Puducherry), PB (Punjab), RJ (Rajasthan), SK (Sikkim), TN (Tamil Nadu), TG (Telangana), TR (Tripura), UP (Uttar Pradesh), UT (Uttarakhand) WB (West Bengal). 

\section{Result and Discussion}

We have considered five different clusters (subgroups) of states namely (i) high risk, (ii) moderate-high, (iii) moderate, (iv) moderate-low, and (v) low-risk states with respect to the daily incidence of COVID-19 positive cases. As such the value of $k$ is taken as five in the incrementalKMN method. The detailed analysis of COVID-19 data set phases wise are narrated in the next 6 subsections.

\subsection{Scenario of novel coronavirus(COVID-19) in India Before Lockdown:}
\begin{figure}
	\centering
	\includegraphics[width=11cm,height=7cm,keepaspectratio]{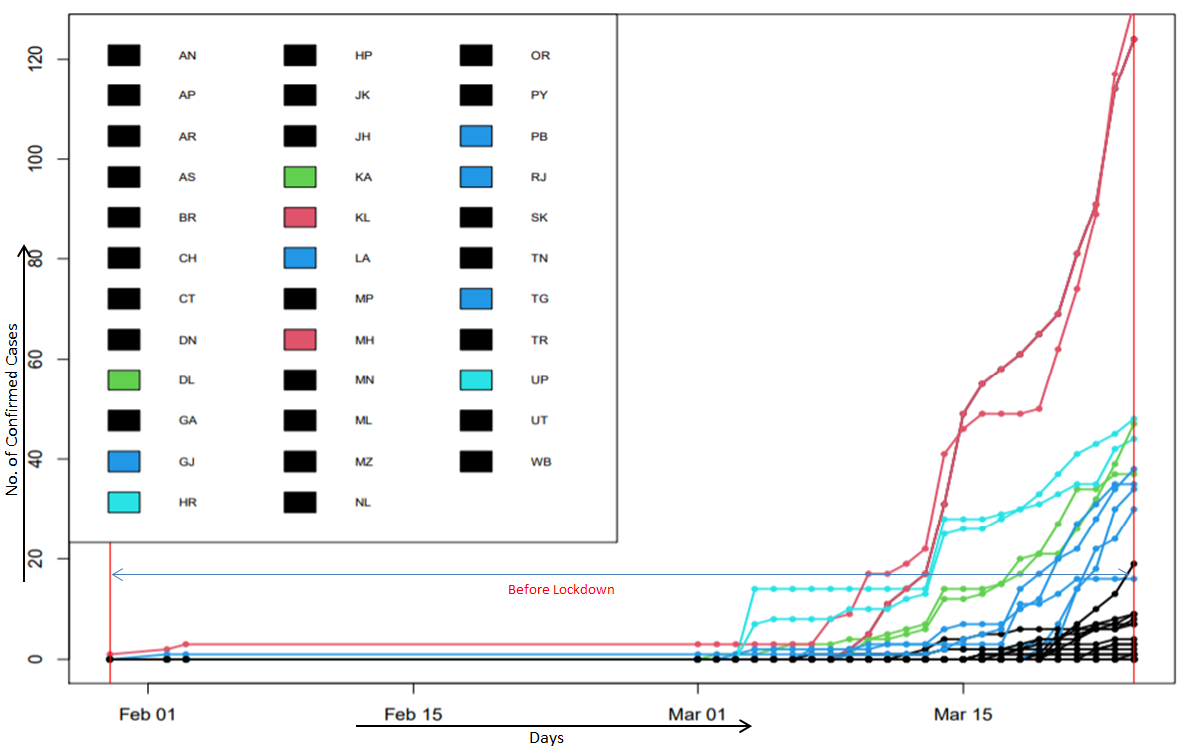}
	\caption{Result of COVID-19 dataset befor lockdown usiung IncrementalKMN clustering method(\textit{k}=5)}
	\label{fig1}
\end{figure}

The first confirmed case of novel coronavirus in India was reported on $30^{th}$ January 2020 in the state of Kerala (Ward, A. (2020, March 24)). The number of confirmed COVID-19 positive cases reached close to 500 on $19^{th}$ March 2020 and by $24^{th}$ March 2020 that is before lockdown the positive case reached 658. In Fig \ref{fig1} which depicts the situation prior to the start of the lockdown based on the number of confirmed cases, the states \{MH and KL\} were on the high-risk state, \{HR and UP\} were on moderate-high risk state, Union Territory DL and the state KA were on moderate risk state, \{GJ, LA, PB, RJ, TG\} were on moderate-low risk states and the remaining states and UTs were on the low-risk states.
\newline

\begin{table*}[htbp]
	\caption{Results on Novel coronavirus(COVID-19) before lockdown Dataset in India}
	\begin{center}
		\begin{tabular}{|c|c|c|c|}
			\hline
			\textbf{\textit{Cluster\_ID}} & \multicolumn{2}{c}{\underline{\textbf{\textit{Before Lockdown }}}} & \textbf{\textit{List of States and UTs}} \\
			&Growth& Doubling&\\
			&Rate&Time(days)&\\
			&(Approx)&&\\
			\hline
			I &	23\% &	3.35 &KL,MH \\
			\hline
			II &	22\% &	3.48 &DL,KA \\
			\hline
			III &	26\% &	3.00 &GJ,LA,PB,RJ,TG\\
			\hline
			IV &	8\% &	9.01 &HR,UP \\
			\hline
			V    &	7\%   &  10.25  &AN,AP,AR,AS,BR,\\
			&&&CH,CT,DN,GA,HP,\\
			&&&JK,JH,MP,MN,ML,\\
			&&&MZ,NL,OR,PY,SK,\\
			&&&TN,TR,UT,WB\\
			\hline
		\end{tabular}
		\label{tab1}
	\end{center}
\end{table*}
In Table \ref{tab1}, the growth rate and doubling time(in days) of each cluster of states are shown. The growth rate of clusters I, II, and III states i.e. \{GJ, LA, PB, RJ, TG\}, \{KL, MH\}, \{DL, KA\} were approximately same and doubling time was 3-4 days (approximate). Whereas, the growth rate of cluster IV i.e. the states \{HR and UP\} were much lees that the top 3 cluster but close to that of cluster V, and doubling time of 9 days(approximately)was nearly 3 time that of the top clusters. The rest of the states and UTs before lockdown, the growth rate was low, and the doubling time was approximately 10-11 Days. 

\subsection{Scenario of Novel coronavirus(COVID-19) in India Lockdown Phase-I:}
\begin{figure}
	\centering
	\includegraphics[width=11cm,height=7cm,keepaspectratio]{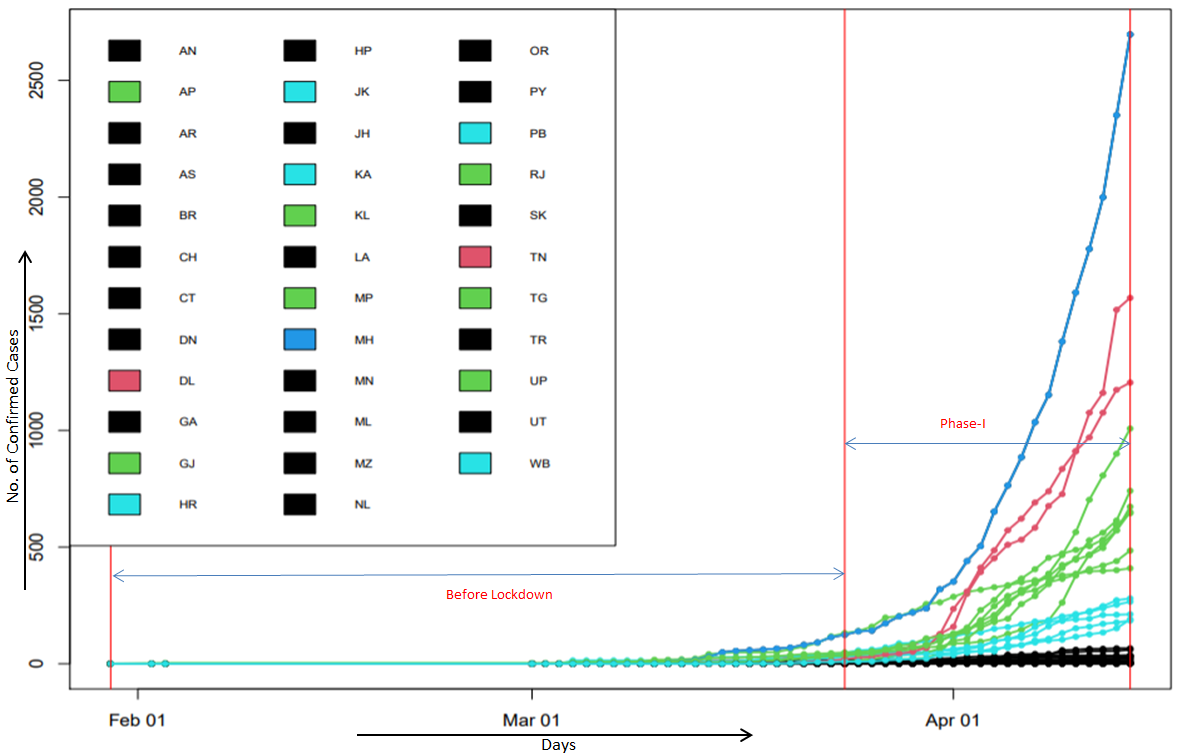}
	\caption{Result of COVID-19 Lockdown Phase-I dataset usiung IncrementalKMN clustering method(\textit{k}=5)}
	\label{fig2}
\end{figure}

The first lockdown in India started on $25^{th}$ March 2020, when nearly all services and factories were suspended and continued till $14^{th}$ April 2020 (Singh, K. D., Goel, V., Kumar, H., Gettleman, J. (2020, March 25)). In Fig \ref{fig2} based on the number of confirmed cases, the state MH was still on the high-risk category state, the Union Territory DL and the state TN were on moderate-high risk state. Whereas the states \{AP, GJ, KL, MP, RJ, TG, UP\} were on moderate and \{HR, JK, KA, PB, WB\} were on moderate-low risk category and followed by the following states and \{UTs AN, AR, AS, BR, CH, CT, DN, GA, HP, JK, JH, LA, MN, ML, MZ, NL, OR, PY, SK, TR, UT\}  on the low-risk category of states.
\newline
\begin{table*}[htbp]
	\caption{Results on Novel coronavirus(COVID-19) Phase-I Dataset in India}
	\begin{center}
		\begin{tabular}{|c|c| c| c|c|c|}
			\hline
			\textbf{\textit{Cluster}} & \multicolumn{2}{c}{\underline{\textbf{\textit{Before Lockdown }}}}&\multicolumn{2}{c}{\underline{\textbf{\textit{Phase-I}}}} & \textbf{\textit{States}} \\
			\textit{\textbf{Id}}&Growth& Doub-&Growth& Doub-&\textbf{\textit{and}}\\
			&Rate&ling&Rate&ling&\textbf{\textit{UTs}}\\
			&(Approx)&Time&(Approx)&Time&\\
			\hline
			I&	29\% &2.72  & 15\%&4.96& MH\\
			\hline
			II&	17\% & 4.41 & 19\%& 3.98 & DL,TN \\
			\hline
			III&	22\% &3.48  &15\% & 4.96& AP,GJ,KL,MP,RJ,TG,UP \\
			\hline
			IV&	20\% &	3.80 &11\% &6.64 &HR,JK,KA,PB,WB\\
			\hline
			V&	4\%  &17.67 & 8\% &  9.01 & AN,AR,AS,BR,CH,CT,DN,\\
			&&&&&GA,HP,JH,LA,MN,ML,\\
			&&&&&MZ,NL,OR,PY,SK,TR,UT\\
			\hline
		\end{tabular}
		\label{tab2}
	\end{center}
\end{table*}
The growth rate and doubling time for all the states in different clusters are shown in Table \ref{tab2}. Accordingly, the growth rate of the state of MH before the lockdown was high and in lockdown phase-I, the growth rate decreased considerably from $29\%$ to $15\%$. Similarly, the 3 days doubling time before the lock down climbed up to 5 days after lockdown phase-I.  On the contrary, the average growth rate of cluster II i.e. \{DL, TN\} has increased, and doubling time in lockdown phase-I was decreased as compared to before lockdown. Similarly, the average growth rate and doubling time of cluster III and IV i.e. states \{AP, GJ, KL, MP, RJ, TG, UP\} and states \{HR, JK, KA, PB, WB\} have improved in lockdown phase-I. But in cluster V i.e. states and UTs \{AN, AR, AS, BR, CH, CT, DN, GA, HP, JH, LA, MN, ML, MZ, NL, OR, PY, SK, TR, UT\}, the average growth rate has increased and doubling time also decreased in lockdown phase-I as compared to before lockdown.  

\subsection{Scenario of Novel coronavirus(COVID-19) in India Lock down Phase-II:}
\begin{figure}
	\centering
	\includegraphics[width=11cm,height=7cm,keepaspectratio]{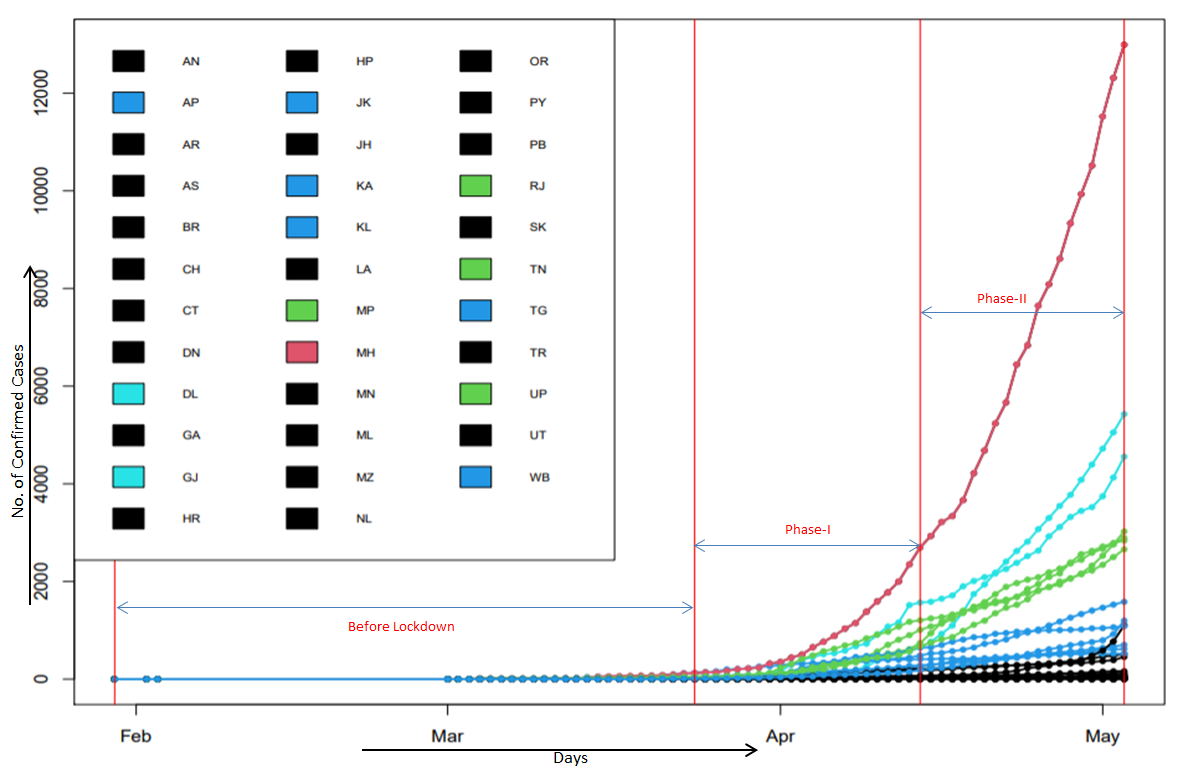}
	\caption{Result of COVID-19 Lock down Phase-II data set using IncrementalKMN clustering method(\textit{k}=5)}
	\label{fig3}
\end{figure}

In phase-II, the lockdown was extended nationwide up to $3^{rd}$ May 2020 with certain relaxations (Bhaskar, U. (2020, April 14), Dutta, P. K. (2020, April 14)). In Fig \ref{fig3} based on the number of confirmed cases, the state MH was again sitting in the high-risk category of states and the Union Territory DL and state GJ were on moderate-high risk state. Similarly, the states \{MP, RJ, TN, UP\} were on moderate risk state and \{AP, JK, KA, KL, TG, WB\} were on moderate-low risk state, and the rest of the states and UTs were in the low-risk state. 
\newline
\begin{table*}[htbp]
	\caption{Results on Novel coronavirus(COVID-19) Phase-II Dataset in India}
	\begin{center}
		\begin{tabular}{|c|c|c|c|c|c|c|c|}
			\hline
			\textbf{\textit{Cluster}} & \multicolumn{2}{c}{\underline{\textbf{\textit{Before Lockdown }}}}&\multicolumn{2}{c}{\underline{\textbf{\textit{Phase-I}}}} &\multicolumn{2}{c}{\underline{\textbf{\textit{Phase-II}}}}& \textbf{\textit{States}} \\
			\textit{\textbf{Id}}&Growth& Doub-&Growth& Doub-&Growth& Doub-&\textbf{\textit{and}}\\
			&Rate&ling&Rate&ling&Rate&ling&\textbf{\textit{UTs}}\\
			&(Approx)&Time&(Approx)&Time&(Approx)&Time&\\
			\hline
			I&29\%&2.72&15\%&4.96&8\%&9.01&MH\\
			\hline
			II&39\%&2.10&17\%&4.41&8\%&9.01&DL,GJ\\
			\hline
			III&14\%&5.29&17\%&4.41&6\%&11.89&MP,RJ,TN,UP\\
			\hline
			IV&21\%&3.64&13\%&5.67&5\%&14.21&AP,JK,KA,KL,TG,WB\\
			\hline
			V&5\%&14.21&8\%&9.01&4\%&17.67&	AN,AR,AS,BR,\\
			&&&&&&&CH,CT,DN,GA,\\
			&&&&&&&HR,HP,JH,LA,\\
			&&&&&&&MN,ML,MZ,NL,OR,\\
			&&&&&&&PY,PB,SK,TR,UT\\
			\hline
		\end{tabular}
		\label{tab3}
	\end{center}
\end{table*}
Table \ref{tab3} shows the growth of the rate and doubling time of each cluster of states and UTs. The growth rate for both the cluster I that is  \{MH\} and cluster II i.e.  \{DL and GJ\} reported as  8\% is a substantial decrease from their 15\% and 17\% in lockdown phase-I respectively. Similarly, the doubling time of approximately 9 days for these two clusters have nearly doubled as compared to lock down phase-I. 
Similarly, the growth rate of cluster III comprising \{MP, RJ, TN, UP\} has decreased from 17\% in lockdown phase -I to 6\% (approximately) in lockdown phase-II, and doubling time has gone from 5 days to 12 days(approximately). 
The growth rate of cluster IV i.e., \{AP, JK, KA, KL, TG, WB\} has decreased from 13\%(approximately) in lockdown phase-I to 5\% in lockdown phase-II, and doubling time has increased from 6 days to 14 days(approximately). 
Finally, the growth rate of cluster V i.e. \{AN, AR, AS, BR, CH, CT, DN, GA, HR, HP, JH, LA, MN, ML, MZ, NL, OR, PY, PB, SK, TR, UT\} has improved in fact nearly halved as compared to lock down phase-I and the doubling time of 18 days(approximately) was nearly double as compared to 9 days lockdown phase-I. 

\subsection{Scenario of Novel coronavirus(COVID-19) in India Lockdown Phase-III:}
\begin{figure}
	\centering
	\includegraphics[width=11cm,height=7cm,keepaspectratio]{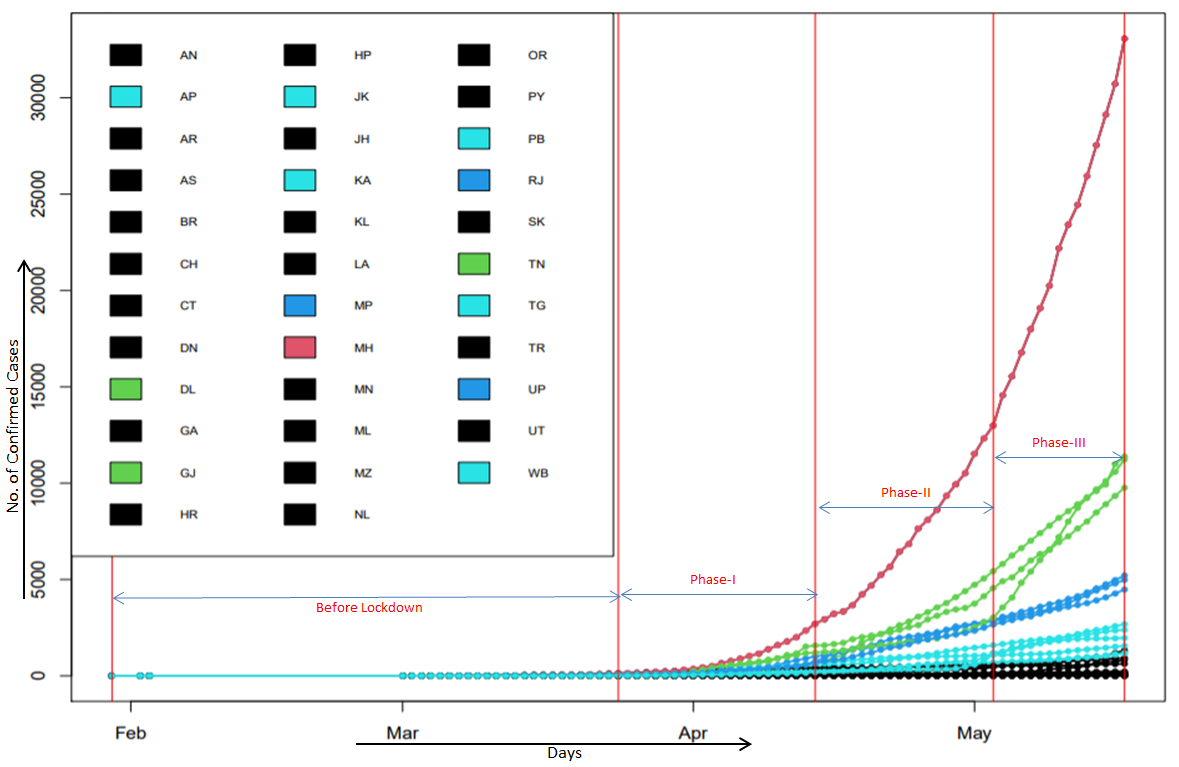}
	\caption{Result of COVID-19 Lockdown Phase-III dataset usiung IncrementalKMN clustering method(\textit{k}=5)}
	\label{fig4}
\end{figure}

Third phase of lockdown started from $4^{th}$–$17^{th}$ May,2020 with some more relaxations ( Online, E. T. (2020, May 4), newsworld24. (2020, May 2)). The country was divided into 3 zones: red zones, orange zones, and green zones (Thacker, T. (2020, May 8)). In Fig \ref{fig4}, again the state MH was in a high-risk category, and DL, GJ, and TN formed the moderate-high risk category. The states \{MP, RJ, UP\} were in moderate risk, \{AP, JK, KA, PB, TG, WB\} were in moderate-low risk state, and the rest of the states and UTs were in the low-risk category.
\newline

\begin{table*}[htbp]
	\caption{Results on Novel coronavirus(COVID-19) Phase-III Dataset in India}
	\begin{center}
	\tiny
		\begin{tabular}{|c|c|c|c|c|c|c|c|c|c|}
			\hline
			\textbf{\textit{Cluster}} & \multicolumn{2}{c}{\underline{\textbf{\textit{Before Lockdown }}}}&\multicolumn{2}{c}{\underline{\textbf{\textit{Phase-I}}}} &\multicolumn{2}{c}{\underline{\textbf{\textit{Phase-II}}}}&\multicolumn{2}{c}{\underline{\textbf{\textit{Phase-III}}}}& \textbf{\textit{States}} \\
			\textit{\textbf{Id}}&Growth& Doub-&Growth& Doub-&Growth& Doub-&Growth& Doub-&\textbf{\textit{and}}\\
			&Rate&ling&Rate&ling&Rate&ling&Rate&ling&\textbf{\textit{UTs}}\\
			&(Approx)&Time&(Approx)&Time&(Approx)&Time&(Approx)&Time&\\
			\hline
			I&29\%&2.72&15\%&4.96&8\%&9.01&6\%&11.89& MH\\
			\hline
			II&32\%&2.50&18\%&4.19&7\%&10.24&6\%&11.89&DL,GJ,TN\\
			\hline
			III&13\%&5.67&17\%&4.41&6\%&11.89&4\%&17.67&MP,RJ,UP\\
			\hline
			IV&22\%&3.48&14\%&5.29&	6\%&11.89&4\%&17.67& AP,JK,KA,\\
			&&&&&&&&&PB,TG,WB\\
			\hline
			V&5\%&14.21&8\%&9.01&3\%&23.45&4\%&17.67& AN,AR,AS,\\
			&&&&&&&&&BR,CH,CT,\\
			&&&&&&&&&DN,GA,HR,\\
			&&&&&&&&&HP,JH,KL,\\
			&&&&&&&&&LA,MN,ML,\\
			&&&&&&&&&MZ,NL,OR,\\
			&&&&&&&&&PY,SK,TR,\\
			&&&&&&&&&UT\\
			\hline
		\end{tabular}
		\label{tab4}
	\end{center}
\end{table*}

In phase-III the growth rate of confirmed cases of all clusters except cluster V in lockdown have decreased and the doubling time of all clusters increased except for cluster V which shows in Table \ref{tab4}. The doubling time of cluster I and II i.e. \{MH\} and \{DL, GJ, TN\} have increased to 12 days(approximate). Similarly, the doubling time of cluster III, IV and V i.e. the states and UTs \{MP, RJ, UP\}, \{AP, JK, KA, PB, TG, WB\} and \{AN, AR, AS, BR, CH, CT, DN, GA, HR, HP, JH, KL, LA, MN, ML, MZ, NL, OR, PY, SK, TR, UT\} have gone down to 18 days approximately.

\subsection{Scenario of Novel coronavirus(COVID-19) in India Lockdown Phase-IV:}
\begin{figure}
	\centering
	\includegraphics[width=11cm,height=7cm,keepaspectratio]{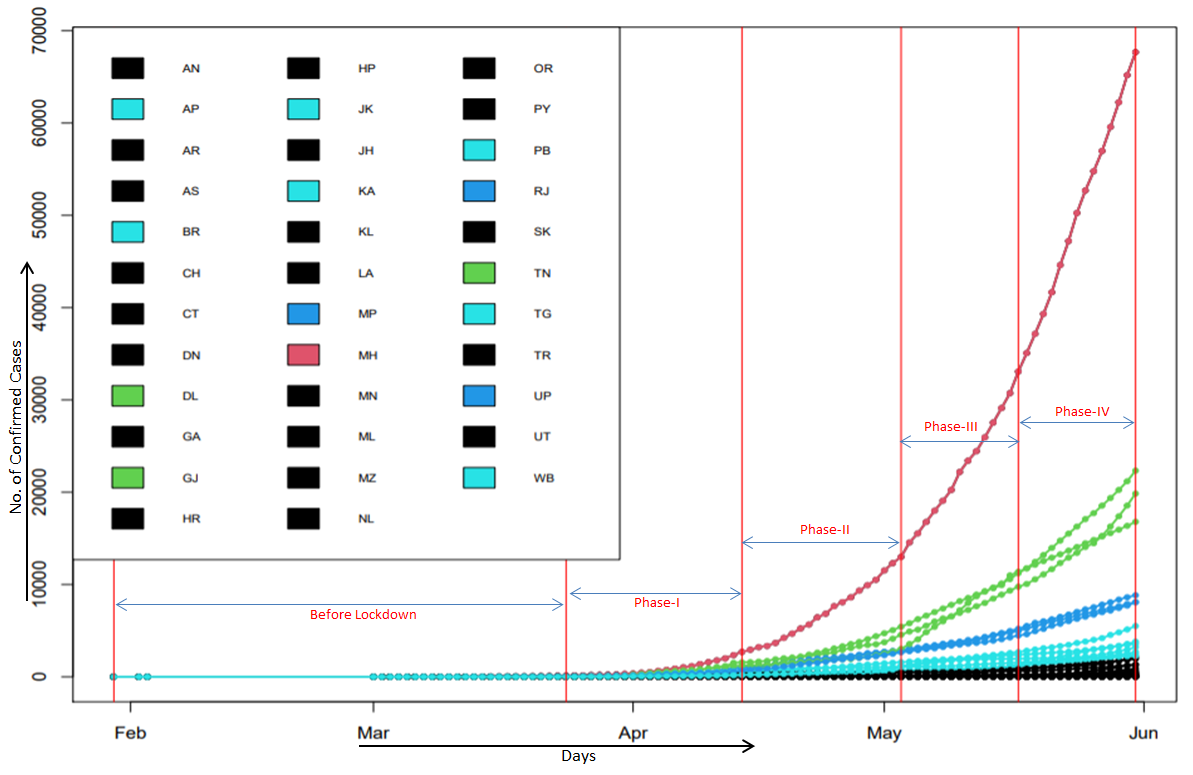}
	\caption{Result of COVID-19 Lockdown Phase-IV dataset using IncrementalKMN clustering method(\textit{k}=5)}
	\label{fig5}
\end{figure}

In phase-IV, the lockdown was extended for another two weeks from $17^{th}-31^{th}$ May 2020 with some additional relaxations. Here, red zones were further divided into to containment and buffer zones (Banerjea, W. B. A. (2020, May 17) , Desk, T. H. N. (2020, May 18) , Online, E. T. (2020, May 21)). In Fig \ref{fig5}, the state MH was remained in the high-risk state based on the number of confirmed cases, and \{DL, GJ, TN\} were in moderate-high risk state. The states \{MP, RJ\} were in a moderate-risk group, \{AP, BR, JK, KA, PB, TG, WB\} were in moderate-low risk, and the rest of the states were in the low-risk state.
\newline	
The growth rate of clusters I and II decreased marginally while that of II and IV remained almost the same as in the previous phases of lockdown. Now the doubling time of cluster I  i.e. the state  MH is 14 days(approximate) and cluster II, III, and IV i.e. the states and UTs \{DL, GJ, TN\}, \{MP, RJ, UP\}, and \{AP, BR, JK, KA, PB, TG, WB\} are all 18 days(approximate). But for cluster V the growth rate more than doubled from phase II. This may be attributed to the homecoming of migrants from the red zone. Similarly, the situation is reflected with the doubling time as well. Except for the cluster V, all clusters in lockdown phase-IV has increased their doubling time. For cluster V,i.e. the states and UTs  \{AN, AR, AS, CH, CT, DN, GA, HR, HP, JH, KL, LA, MN, ML, MZ, NL, OR, PY, SK, TR, UT\}  doubly time deteriorated from 18 days to 8 days in Table \ref{tab5}. 

\subsection{Scenario of Novel coronavirus(COVID-19) in India Lockdown Phase-V:}
\begin{figure}
	\centering
	\includegraphics[width=11cm,height=7cm,keepaspectratio]{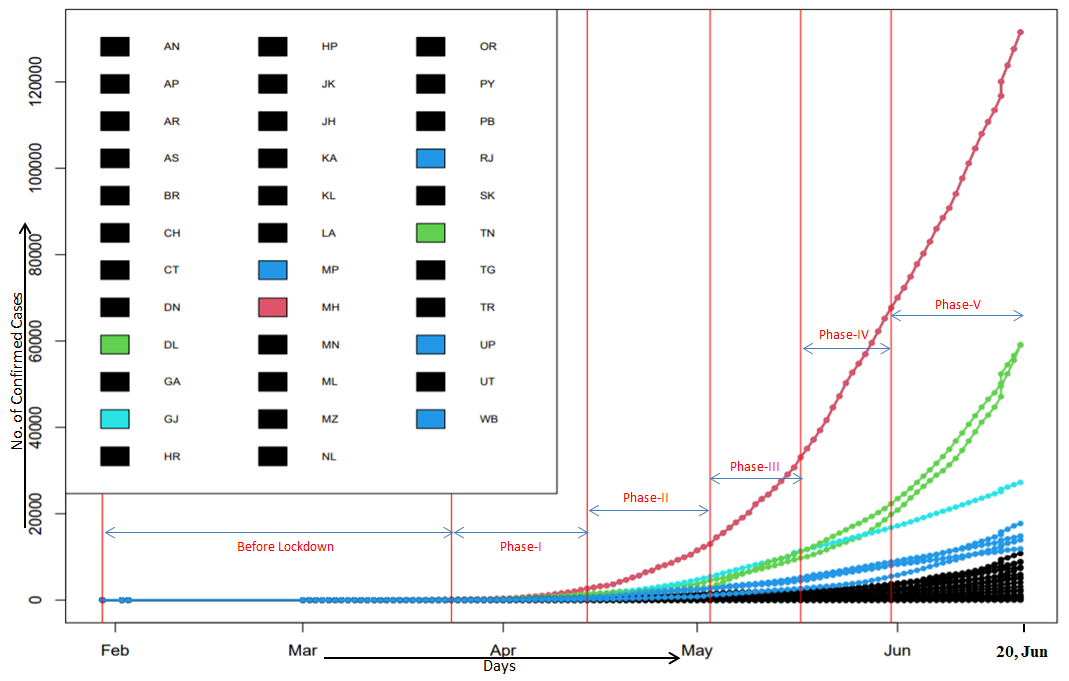}
	\caption{Result of COVID-19 Lockdown Phase-V dataset usiung IncrementalKMN clustering method(\textit{k}=5)}
	\label{fig6}
\end{figure}

\begin{figure}
	\centering
	\includegraphics[width=11cm,height=7cm,keepaspectratio]{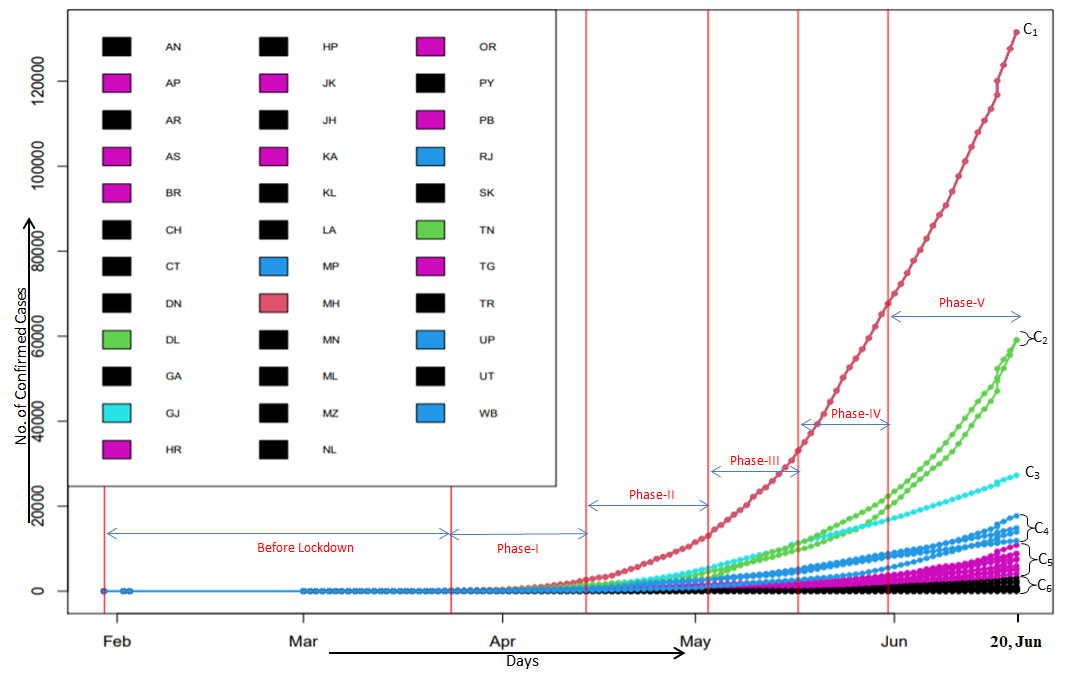}
	\caption{Result of COVID-19 Lockdown Phase-V dataset usiung IncrementalKMN clustering method(\textit{k}=6)}
	\label{fig8}
\end{figure}

\begin{figure}
	\centering
	\includegraphics[width=11cm,height=7cm,keepaspectratio]{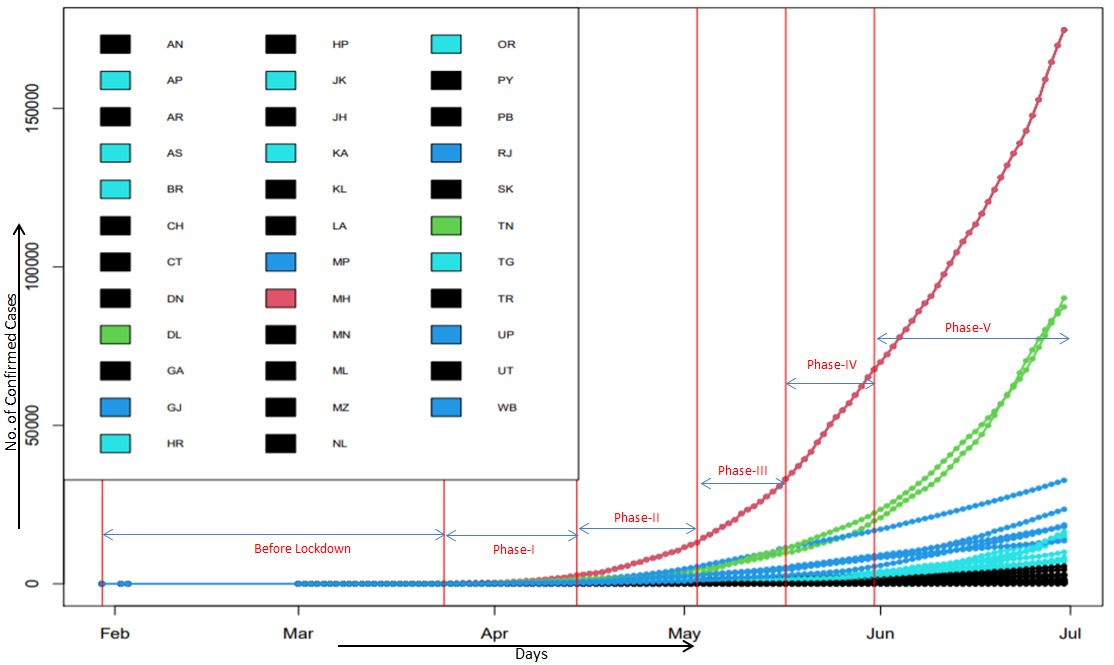}
	\caption{Result of COVID-19 Lockdown Phase-V dataset usiung IncrementalKMN clustering method(\textit{k}=5)}
	\label{fig9}
\end{figure}

The phase V (or Unlock-I) of lockdown started from $1^{th}-30^{th}$ June 2020 with only limited restrictions (Sharma, N., Ghosh, D. (2020, May 30)) . In our study, we have considered the novel coronavirus (COVID-19) data set till $20^{th}$ June 2020. The state MH was in the high-risk state based on the number of confirmed cases and \{DL, TN\} was in moderate-high risk state which shows in figure \ref{fig6}. Similarly, the state GJ was in moderate risk state, and \{MP, RJ, UP, WB\} were in moderate-low risk state, and the rest of the states and UTs were in the low-risk state. 
\newline
From Table \ref{tab6}, the growth rate of all clusters has decreased or remained the same in lockdown phase-V as compared to previous phases. Similarly, the doubling time of all clusters has increased or remained the same. The doubling time of cluster I i.e. the state MH is in 24 days(approximate). Similarly, cluster II i.e. the states \{DL, TN\} is required 14-15 days(approximate) to double and cluster III i.e. the state GJ is required 35 days to double. The cluster IV and V have  24 days(approximate) and 9 days(approximate) to doubling time.
From what we have found it was expected that some of the states/Uts lying in the Category V  will soon move to the Category IV in the next weak or so starting $21^{th}$ June 2020.
\newline
In order to verify this, we have then extended our study covering data up to $30^{th}$ June 2020 to cover the full Unlock I period to see how the cluster changed in the last 10 days of this phase. The result is shown in figure \ref{fig9}. As expected the some of the states \{AP, AS, BR, HR, JK, KA, OR, PB, TG\} which were in Category V moved to category IV. In fact from figure \ref{fig8} where we have demarcated 6 instead of 5 clusters to reveal hidden groups with the cluster V gave a clear indication of the tendency of the above state towards the next higher risk category (see also Table \ref{tab7}).

\section{Summary of Findings \& limitations}
\begin{enumerate}[i.]
	\item From the high-risk category in the pre lockdown period KL(Kerala) move to the low-risk category while MH(Maharashtra) remained in the high-risk category throughout the period. Whether this can be interpreted as successful implementing of lockdown rules in KL is though arguable.
	\item Lockdown seemingly had its impact on two other states HR(Haryana) and UP(Uttar Pradesh) as these states improved down to low risk and moderate low-risk category. 
	\item Based on that we observed that some of the states like AP (Andhra Pradesh), AS (Assan), BR (Bihar), HR (Haryana), JK (Jammu and Kashmir), KA (Karnataka), OR (Odisha), PB (Punjab), and TG (Telangana) will move up the ladder towards higher risk level in weeks to come.
	\item Also it appears that some of the states/UTs where the onset was late might show surge in coming days to move to a higher category of risk.
	\item The current study is based on the number of confirmed cases and subject to reporting biases if any in the data source. We have not considered other relevant factors, co-variate and non-pharmaceutical interventions which might completely alter the picture favorably.
	\item Number of confirmed cases alone can not give a true picture of the prevalence of the disease as it is proportional to the number of tests conducted.
\end{enumerate}

\section{Conclusion }
In this study, we used incrementalKMN (Prasad, R. K., Sarmah, R., Chakraborty, S.,(2019)) clustering method to classify the Indian states and UTs in \textit{five} different stages of risk on the basis of the number of confirmed cases of novel coronavirus (COVID-19). Then evaluated the growth rate and doubling time of confirmed cases of each cluster. As on $30^{th}$ June 2020, the state MH (Maharastra) in on high-risk category with the doubling time of confirmed cases is 23-24 days(approximate). Similarly, the union territory of DL (Delhi) and the state TN (Tamil Nadu) are in moderate-high risk state and is expected to join MH in the high-risk state soon as the doubling time of these set of states is 14-15 days(approximate). The state GJ(Gujarat) is in moderate risk sate and has decreased their growth rate during the lockdown phases and the doubling time of this state is 35 days(approximate). Moreover, the states MP (Madhya Pradesh), RJ (Rajasthan), UP (Uttar Pradesh), and WB (West Bengal) are in moderate-low risk state and the growth rate and doubling time are same as the state of MH. Rest of the states are in the low-risk cluster but some of the states namely AP (Andhra Pradesh), AS (Assam), BR (Bihar), HR (Haryana), JK (Jammu and Kashmir), KA (Karnataka), OR (Odisha), PB (Punjab), and the TG (Telangana) are expected to reach next level of risk soon. Our aim in this work is not to prove or disprove anything but present the pattern for everyone to see and realize that the situation we are in still remains critical. There is no room to rejoice now by saying we are in low risk compared to MH. Yes! MH is ahead of us but not in terms of numbers but we are behind them in time scale and it is a matter of time before we reach and experience that stage.

\begin{landscape}
	\begin{table*}[htbp]
		\caption{Results on Results on Novel coronavirus(COVID-19) Phase-IV Dataset in India}
		\begin{center}
			\small
			\begin{tabular}{|c|c|c|c|c|c|c|c|c|c|c|c|}
				\hline
				\textbf{\textit{Cluster}} & \multicolumn{2}{c}{\underline{\textbf{\textit{Before Lockdown }}}}&\multicolumn{2}{c}{\underline{\textbf{\textit{Phase-I}}}} &\multicolumn{2}{c}{\underline{\textbf{\textit{Phase-II}}}}&\multicolumn{2}{c}{\underline{\textbf{\textit{Phase-III}}}}&\multicolumn{2}{c}{\underline{\textbf{\textit{Phase-IV}}}} & \textbf{\textit{States}} \\
				\textit{\textbf{Id}}&Growth& Doub-&Growth& Doub-&Growth& Doub-&Growth& Doub-&Growth& Doub-&\textbf{\textit{and}}\\
				&Rate&ling&Rate&ling&Rate&ling&Rate&ling&Rate&ling&\textbf{\textit{UTs}}\\
				&(Approx)&Time&(Approx)&Time&(Approx)&Time&(Approx)&Time&(Approx)&Time&\\
				\hline
				I&29\%&2.72&15\%&4.95&8\%&9.01&6\%&11.89&5\%&14.21&MH\\
				\hline
				II&32\%&2.50&18\%&4.19&7\%&10.24&6\%&11.89&4\%&17.67& DL,GJ,TN\\
				\hline
				III&13\%&5.67&17\%&4.414&6\%&11.89&4\%&17.67&4\%&17.67&MP,RJ,UP\\
				\hline
				IV&21\%&3.63&14\%&5.29&7\%&10.24&4\%&17.67&4\%&17.67&AP,BR,JK,KA,\\
				&&&&&&&&&&&PB, TG, WB\\
				\hline
				V&5\%&14.21 &8\%&9.01&3\%&23.45&4\%&17.67&9\%&8.04& AN,AR,AS,CH, \\
				&&&&&&&&&&&CT,DN,GA,HR,\\
				&&&&&&&&&&&HP,JH,KL,LA,\\
				&&&&&&&&&&&MN,ML,MZ,NL,\\
				&&&&&&&&&&&OR,PY,SK,TR,UT\\
				\hline
			\end{tabular}
			\label{tab5}
		\end{center}
	\end{table*}
\end{landscape}
\begin{landscape}
	\begin{table*}[htbp]
		\caption{Results on Novel coronavirus(COVID-19) Phase-V Dataset till 20, June 2020 in India}
		\begin{center}
			\small
			\begin{tabular}{|c|c|c|c|c|c|c|c|c|c|c|c|c|c|}
				\hline
				\textbf{\textit{Cluster}} & \multicolumn{2}{c}{\underline{\textbf{\textit{Before Lockdown }}}}&\multicolumn{2}{c}{\underline{\textbf{\textit{Phase-I}}}} &\multicolumn{2}{c}{\underline{\textbf{\textit{Phase-II}}}}&\multicolumn{2}{c}{\underline{\textbf{\textit{Phase-III}}}}&\multicolumn{2}{c}{\underline{\textbf{\textit{Phase-IV}}}}&\multicolumn{2}{c}{\underline{\textbf{\textit{Phase-V till 20, June}}}} & \textbf{\textit{States}} \\
				\textit{\textbf{Id}}&Growth& Doub-&Growth& Doub-&Growth& Doub-&Growth& Doub-&Growth& Doub-&Growth& Doub-&\textbf{\textit{and}}\\
				&Rate&ling&Rate&ling&Rate&ling&Rate&ling&Rate&ling&Rate&ling&\textbf{\textit{UTs}}\\
				&(Approx)&Time&(Approx)&Time&(Approx)&Time&(Approx)&Time&(Approx)&Time&(Approx)&Time&\\
				\hline
				I&29\% &2.72 &15\% &4.96 &8\%&9.01 &6\%&11.89 &5\%&14.21&3\%&23.45&MH \\
				\hline
				II&17\%&4.41 &19\%&3.98 &5\% &14.21&7\% &10.24&5\% &14.21&5\%&14.21 &DL,TN\\
				\hline
				III&60\%&1.47&14\%&5.29&11\%&6.64&5\%&14.21&3\%&23.45&2\%&35.00&GJ\\
				\hline
				IV&18\%&4.19&16\%&4.67&7\%&10.24&4\%&17.67&4\%&17.67&3\%&23.45&MP,RJ,\\
				&&&&&&&&&&&&&UP,WB\\
				\hline
				V& 8\% &9.01& 9\% &8.04 & 4\%&17.67& 4\%&17.67 &8\%&9.01& 8\%	& 9.01&AN,AP,AR,\\
				&&&&&&&&&&&&&AS,BR,CH,\\
				&&&&&&&&&&&&&CT,DN,GA,\\
				&&&&&&&&&&&&&HR,HP,JK,\\
				&&&&&&&&&&&&&JH,KA,KL,\\
				&&&&&&&&&&&&&LA,MN,ML,\\
				&&&&&&&&&&&&&MZ,NL,OR,\\
				&&&&&&&&&&&&&PB,PY,SK,\\
				&&&&&&&&&&&&&TG,TR,UT.\\
				\hline
			\end{tabular}
			\label{tab6}
		\end{center}
	\end{table*}
\end{landscape}
\begin{landscape}

	\begin{table*}[htbp]
		\caption{Results on Novel coronavirus(COVID-19) Phase-V Dataset in India}
		\begin{center}
			\small
			\begin{tabular}{|c|c|c|c|c|c|c|c|c|c|c|c|c|c|}
				\hline
				\textbf{\textit{Cluster}} & \multicolumn{2}{c}{\underline{\textbf{\textit{Before Lockdown }}}}&\multicolumn{2}{c}{\underline{\textbf{\textit{Phase-I}}}} &\multicolumn{2}{c}{\underline{\textbf{\textit{Phase-II}}}}&\multicolumn{2}{c}{\underline{\textbf{\textit{Phase-III}}}}&\multicolumn{2}{c}{\underline{\textbf{\textit{Phase-IV}}}}&\multicolumn{2}{c}{\underline{\textbf{\textit{Phase-V }}}} & \textbf{\textit{States}} \\
				\textit{\textbf{Id}}&Growth& Doub-&Growth& Doub-&Growth& Doub-&Growth& Doub-&Growth& Doub-&Growth& Doub-&\textbf{\textit{and}}\\
				&Rate&ling&Rate&ling&Rate&ling&Rate&ling&Rate&ling&Rate&ling&\textbf{\textit{UTs}}\\
				&(Approx)&Time&(Approx)&Time&(Approx)&Time&(Approx)&Time&(Approx)&Time&(Approx)&Time&\\
				\hline
				I&29\% &2.72 &15\% &4.96 &8\%&9.01 &6\%&11.89 &5\%&14.21&3\%&23.45&MH \\
				\hline
				II&17\%&4.41 &19\%&3.98 &5\% &14.21&7\% &10.24&5\% &14.21&5\%&14.21 &DL,TN\\
				\hline
				
				III &	26\% & 3.00 &	16\% & 4.67 &	8\% & 9.01 &	4\% & 17.67 &	4\% & 17.67 &	3\% & 23.45 & GJ,MP,RJ,UP,\\
				&&&&&&&&&&&&& WB.\\
				\hline
				IV &	12\% & 6.12	& 15\% & 4.96 & 5\% & 14.21 &	5\% & 14.21 &	7\% & 10.24 &	5\% & 14.21 & AP,AS,BR,HR,\\
				&&&&&&&&&&&&& JK,KA,OR,TG\\
				\hline
				V & 6\% & 11.89 &	6\% & 11.89 &	3\% &23.45&	3\% & 23.45 &	9\% & 8.04 &	7\% & 10.24 & AN,AR,CH,CT,\\
				&&&&&&&&&&&&& DN,GA,HP,JH,\\
				&&&&&&&&&&&&& KL,LA,MN,ML,\\
				&&&&&&&&&&&&& MZ,NL,PY,PB,\\
				&&&&&&&&&&&&& SK,TR,UT\\
				\hline
			\end{tabular}
			\label{tab7}
		\end{center}
	\end{table*}
\end{landscape}

\section*{References}
Prasad, R. K., Sarmah, R., Chakraborty, S. (2019). Incremental k-Means Method. Lecture Notes in Computer Science Pattern Recognition and Machine Intelligence, 38–46.\\
\\
Ward, A. (2020, March 24). India's coronavirus lockdown and its looming crisis, explained. Vox. Vox. Retrieved June 23, 2020, from https://www.vox.com/2020/3/24/\\
21190868/coronavirus-india-modi-lockdown-kashmir\\
\\	
Singh, K. D., Goel, V., Kumar, H., Gettleman, J. (2020, March 25). India, Day 1: World's Largest Coronavirus Lockdown Begins. Retrieved June 23, 2020, from https://www.\\
nytimes.com/2020/03/25/world/asia/india-lockdown-coronavirus.html.\\
\\
Bhaskar, U. (2020, April 14). India to remain closed till 3 May, economy to open up gradually in lockdown 2.0. Retrieved June 23, 2020, from https://www.livemint.com/news/india/\\
pm-modi-announces-extension-of-lockdown-till-3-may-11586839412073.html\\
\\	
Dutta, P. K. (2020, April 14). In coronavirus lockdown extension, Modi wields stick, offers carrot on exit route. Retrieved June 23, 2020, from https://www.indiatoday.in/coronavirus-outbreak/story/in-coronavirus-lockdown-extension-modi-wields-stick-offers-carrot-on-exit-route-1666741-2020-04-14\\
\\	
India coronavirus: All major cities named Covid-19 'red zone' hotspots. (2020, April 16). . Retrieved June 23, 2020, from https://www.bbc.com/news/world-asia-india-52306225\\
\\	
Online, E. T. (2020, May 4). Lockdown extended by 2 weeks, India split into red, green and orange zones. Retrieved June 23, 2020, from https://economictimes.indiatimes.com/\\
news/politics-and-nation/govt-extends-lockdown-by-two-weeks-permits-
considerable-\\
relaxations-in-green-and-orange-zones/articleshow/75491935.cms\\
\\	
newsworld24. (2020, May 2). Lockdown Extension till May 17: Read MHA guidelines. Retrieved June 23, 2020, from https://www.newsworld24.in/2020/05/lockdown-extension-till-may-17-read-mha-guidelines.html\\
\\	
Thacker, T. (2020, May 8). Centre issues state-wise division of Covid-19 red, orange green zones. Retrieved June 23, 2020, from https://economictimes.indiatimes.com/news/politics-and-nation/centre-issues-state-wise-division-of-covid-19-red-orange-green-zones/\\
articleshow/75486277.cms\\
\\	
Banerjea, W. B. A. (2020, May 17). Coronavirus lockdown extended till 31 May, says NDMA. Retrieved June 23, 2020, from https://www.livemint.com/news/india/covid-19-lockdown-4-0-coronavirus-lockdown-extended-till-31-may-says-ndma-11589715203633.html\\
\\	
Desk, T. H. N. (2020, May 18). India lockdown 4.0 guidelines: What's allowed and what's not? Retrieved June 23, 2020, from https://www.thehindu.com/news/national/lockdown-40-guidelines-whats-allowed-and-whats-not/article31609394.ece\\
\\	
Online, E. T. (2020, May 21). Lockdown 4.0 guidelines: Nationwide lockdown extended till May 31, with considerable relaxations. Retrieved June 23, 2020, from https://\\
economictimes.indiatimes.com/news/politics-and-nation/centre-extends-nationwide-\\
lockdown-till-may-31-with-considerable-relaxations/articleshow/75790821.cms\\
\\	
Sharma, N., Ghosh, D. (2020, May 30). "Unlock1": Malls, Restaurants, Places Of Worship To Reopen June 8. Retrieved June 23, 2020, from https://www.ndtv.com/india-news/lockdown-extended-till-june-30-malls-restaurants-can-reopen-from-june-8-except-in-containment-zones-2237910\\
\\	
Han, Jiawei, Jian Pei, and Micheline Kamber. Data mining: concepts and techniques. Elsevier, 2011.\\
\\	
Murphy, C. B. (2020, May 15). Understanding the Compound Annual Growth Rate – CAGR. Retrieved June 25, 2020, from https://www.investopedia.com/terms/c/cagr.asp\\
\\	
Manias, M. (2020, January 10). Doubling Time Calculator. Retrieved June 25, 2020, from https://www.omnicalculator.com/math/doubling\_time\\
\\		
Zarikas V, Poulopoulos SG, Gareiou Z, Zervas E (2020) Clustering analysis of countries using the COVID-19 cases dataset. Data in Brief 31:105787. doi: 10.1016/j.dib.2020.105787\\
\\				
Kumar S (2020) Monitoring Novel Corona Virus (COVID-19) Infections in India by Cluster Analysis. Annals of Data Science. doi: 10.1007/s40745-020-00289-7\\
\\			
Kumar A, Gupta PK, Srivastava A (2020) A review of modern technologies for tackling COVID-19 pandemic. Diabete \& Metabolic Syndrome: Clinical Research \& Reviews 14:569–573. doi: 10.1016/j.dsx.2020.05.008\\
\\			
Blumenstock J (2020) Machine learning can help get COVID-19 aid to those who need it most. Nature. doi: 10.1038/d41586-020-01393-7\\
\\			
Ozturk T, Talo M, Yildirim EA, et al (2020) Automated detection of COVID-19 cases using deep neural networks with X-ray images. Computers in Biology and Medicine 121:103792. doi: 10.1016/j.compbiomed.2020.103792\\
\\			
Javaid M, Haleem A, Vaishya R, et al (2020) Industry 4.0 technologies and their applications in fighting COVID-19 pandemic. Diabetes \& Metabolic Syndrome: Clinical Research \& Reviews 14:419–422. doi: 10.1016/j.dsx.2020.04.032\\
\\			
Hassanien AE, Mahdy LN, Ezzat KA, et al (2020) Automatic X-ray COVID-19 Lung Image Classification System based on Multi-Level Thresholding and Support Vector Machine. doi: 10.1101/2020.03.30.20047787
		
\end{document}